\documentclass[fleqn,usenatbib,useAMS]{mnras}
\usepackage{savesym}
\savesymbol{tablenum}
\usepackage{siunitx}
\usepackage{multirow}
\restoresymbol{SIX}{tablenum}
\usepackage{graphicx,url,amssymb,amsmath,color,units,wasysym,epsfig,epstopdf,xcolor,soul,xcolor}
\usepackage{verbatim}
\usepackage{empheq}
\usepackage[normalem]{ulem}
\usepackage{amsmath}
\usepackage{xspace}

\usepackage{txfonts}

\pubyear{ -- }

\begin{document}
\label{firstpage}
\pagerange{\pageref{firstpage}--\pageref{lastpage}}

\title[An asymmetrical model for Cas A ]{An Asymmetrical Model for High Energy Radiation of Cassiopeia A}
\author[Zhan et al.]{Shihong Zhan$^{1,2}$, Wei Wang$^{1,2}$\thanks{E-mail: wangwei2017@whu.edu.cn}, Guobin Mou$^{1,2}$\thanks{gbmou@whu.edu.cn}, Zhuo Li$^{3,4}$\thanks{zhuo.li@pku.edu.cn} \\
$^{1}$School of Physics and Technology, Wuhan University, Wuhan 430072, China\\
$^{2}$WHU-NAOC Joint Center for Astronomy,Wuhan University, Wuhan 430072,China \\
$^{3}$Department of Astronomy, School of Physics, Peking University, Beijing 100871, China \\
$^{4}$Kavli Institute for Astronomy and Astrophysics, Peking University, Beijing 100871, China
}
\maketitle
\begin{abstract}
{Cassiopeia A (Cas A) supernova remnant shows strong radiation from radio to gamma-ray bands. The mechanism of gamma-ray radiation in Cas A and its possible contribution to PeV cosmic rays are still under debate. The X-ray imaging reveals an asymmetric profile of Cas A, suggesting the existence of a jet-like structure. In this work, we propose an asymmetrical model for Cas A, consisting of a fast moving jet-like structure and a slowly expanding isotropic shell. This model can account for the multi-wavelength spectra of Cas A, especially for the power-law hard X-ray spectrum from $\sim$ 60 to 220 keV. The GeV to TeV emission from Cas A should be contributed by both hadronic and leptonic processes. Moreover, the jet-like structure may produce a gamma-ray flux of $\sim 10^{-13}\rm erg\ cm^{-2}\ s^{-1}$ at $\sim 100$ TeV, to be examined by LHAASO and CTA.  }
\end{abstract}


\section{Introduction}
Supernova remnants (SNRs) are widely believed to be one of the candidate accelerators of cosmic rays (CRs) up to TeV or even PeV energy levels. The diffusive shock acceleration (DSA) mechanism \citep{1977DoSSR.234.1306K,1978MNRAS.182..147B,1987PhR...154....1B} in SNR shocks is expected to produce non-thermal particles following an energy distribution of a power law. Given the supernova explosion rate of $\sim$3 per century in the Milky Way, a fraction 5-10\% of the SNR kinetic energy converted into CRs \citep{2012ApJ...755..106P,2016ApJ...821L..20P} can well account for the observed CRs. The radiation mechanism for GeV-TeV gamma-rays from SNRs is still under debate - it could be the leptonic processes, e.g., inverse Compton (IC) scattering or non-thermal bremsstrahlung (NTB) by cosmic ray electrons (CRe), or the hadronic processes via decay of $\pi^{0}$ which are created in proton-proton (pp) collisions. The high-resolution gamma-ray observations combined with the broadband modeling may help to solve the problem of radiation mechanism.

Cassiopeia A, a young (of age $t_{\rm age}\approx 350 \ \rm yr$) SNR with abundant multi-wavelength observations, has long been considered as an very high-energy particle accelerator \citep{2017hsn..book..161K}. Cas A originated from a core-collapse Type IIb supernova explosion \citep{2008Sci...320.1195K} and the progenitor was possibly a red supergiant \citep{2003ApJ...593L..23C}. The distance of Cas A is estimated to be 3.4$\pm 0.4$ kpc \citep{1995ApJ...440..706R}.

As a rare Galactic SNR, Cas A is bright in a broad spectrum from radio to $\gamma$-ray band. The radio observation of Cas A shows a bright ring with a radius of $\sim$1.7 pc \citep{1975Natur.257..463B,1987Natur.327..395B,1995ApJ...455L..59K}, and a faint plateau radio emission with a radius of $\sim$2.5 pc \citep{2014ApJ...785....7D} suggests that there may exist a reverse shock in Cas A. The SNR shock swept-up material is estimated to be compatible with the ejected gas, so the shock of Cas A could be in an intermediate evolution phase between the ejecta-dominated and Sedov phases \citep{1977ARA&A..15..175C,2011hea..book.....L}. 

The X-ray observations of Cas A in 4-6 keV band show rim structure of the forward shock wave \citep{2001ApJ...552L..39G}. The non-thermal X-ray component in the forward shock can be explained by synchrotron radiation of CRe with energy of $\sim$40-60 TeV \citep{2003ApJ...584..758V}. Non-thermal hard X-ray emission from Cas A was also reported from OSSE (40 - 120 keV, \citealt{1996A&AS..120C.357T}), RXTE (2 - 60 keV, \citealt{1997ApJ...487L..97A}) and Suzaku observations \citep{2009PASJ...61.1217M}. The emission of Cas A above 10 keV can be fitted by a power-law model with a power-law index $\Gamma\sim 3.04$ \citep{1997ApJ...487L..97A}. NuSTAR reveals several hot spots in the interior of the Cas A above 15 keV \citep{2015ApJ...802...15G}. Based on the ten-year data of INTEGRAL, \cite{2016ApJ...825..102W} showed the non-thermal hard X-ray emission with a power-law index of $\Gamma \sim 3.1$, which does not show any sign of cutoff even up to $\sim 220$ keV.

The non-thermal X-ray features can reveal the characteristics of the relativistic electrons, while the origin of $\gamma$-rays is quite controversial. The $\gamma$-ray spectrum follows a power-law index of $1.9-2.4$ in GeV - sub TeV band \citep{2010ApJ...710L..92A,2013ApJ...766L...8Y,2014A&A...563A..88S}, and the spectrum exhibits a break at $\sim1.7$ GeV \citep{2013ApJ...766L...8Y,2017MNRAS.472.2956A}. \cite{2001A&A...370..112A} made a first detection in 1-10 TeV band by the HEGRA, with the photon index of $\Gamma\sim 2.5$, which was confirmed by MAGIC \citep{2007A&A...474..937A,2017MNRAS.472.2956A}  and VERITAS \citep{2008AIPC.1085..357H,2010ApJ...714..163A,2015ICRC...34..760K,2020ApJ...894...51A}. MAGIC also reported a spectral cut-off energy at $\sim 3.5$ TeV \citep{2017MNRAS.472.2956A}, suggesting that Cas A may not be a PeVatron at its present age. Moreover, the gamma-ray spectrum based on recent VERITAS data implied that protons can be accelerated up to $\sim 6$ TeV in Cas A  \citep{2020ApJ...894...51A}.

The emissions from radio to X-ray of Cas A are attributed mainly to the synchrotron radiation. The GeV - TeV gamma-ray spectrum could be produced by either pp collisions, or IC and NTB by relativistic electrons. The one-zone model fails in explaining the multi-wavelength spectrum of Cas A \citep{2017MNRAS.472.2956A}, especially for the hard X-rays in $\sim 60 - 220$ keV.  Given the observational evidence for the reverse shock in Cas A \citep{2001ApJ...552L..39G,2004ApJ...614..727M,2008ApJ...685..988T}, a two-zone model with both the forward and reverse shocks has been adopted for the multi-wavelength emission from Cas A in previous works \citep{2008ApJ...686.1094H,2008ApJ...673..271R,2019ApJ...874...98Z}.


In the previous works on Cas A it is usually assumed that a spherical shock is expanding into the medium with a velocity of $\sim 5000-6000 \rm \ km \ s^{-1}$. However, the X-ray imaging observations suggested an asymmetrical explosion in Cas A \citep{2012ApJ...746..130H,2004ApJ...615L.117H}. The asymmetrical explosion scenario was also supported by the $^{44}$Ti emission line observed in Cas A \citep{2014Natur.506..339G,2016ApJ...825..102W}. In this work, based on observations, we put forward an asymmetrical model for Cas A, including a spherical expanding shock (zone 1) and a jet-like structure (zone 2). The main goal of the model is to explain the broadband spectrum of Cas A from radio to $\gamma$-rays, especially the hard X-ray band, and investigate the particle acceleration in the jet-like structure. In Section 2, we show the details of the model. The results of multi-wavelength spectral fitting are shown in Section 3. Finally we discuss implications of the model and draw conclusions in Section 4. In the appendix, we make a brief calculation to show why we do not consider the contribution of secondary electrons from pp collisions in the spectral fitting.

\section{Asymmetrical Model}

\begin{figure}
\begin{minipage}{0.50\textwidth}
 \includegraphics[width=0.92\textwidth]{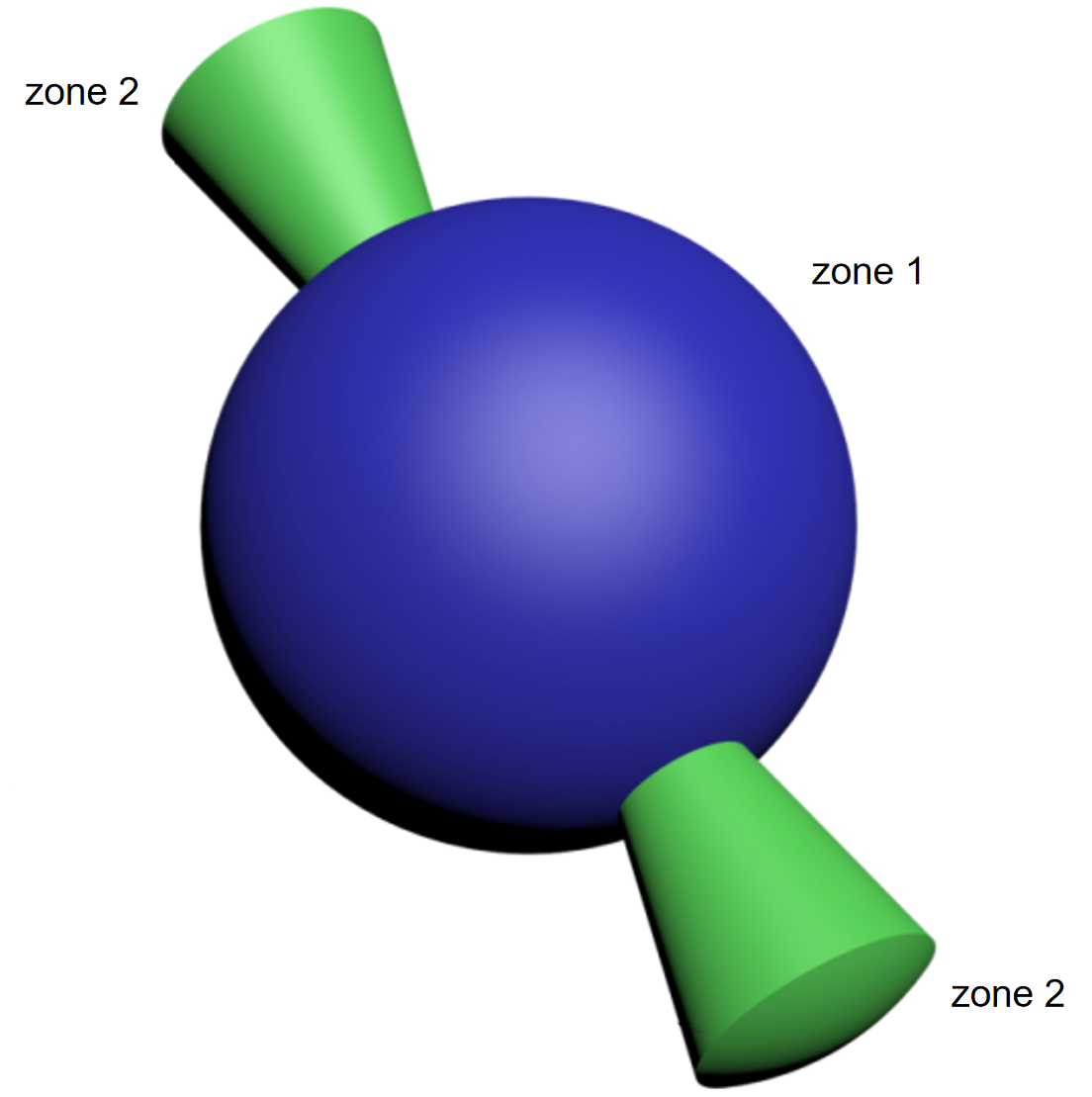}
\end{minipage}
\caption{The sketch of the asymmetrical expansion model. Zone 1 is the isotropically expanding shock wave, and zone 2 is the jet structure driving a fast shock wave. }
\label{fig1}
\end{figure}

According to the X-ray observations \citep{2004ApJ...615L.117H,2006ApJ...645..283F,2011ApJ...732....3R,2014Natur.506..339G,2016ApJ...825..102W}, the explosion of Cas A should not be isotropic. Here we propose an asymmetrical explosion model, as shown in Fig. 1, an isotropic expansion component with a normal velocity, and a jet-like structure with a higher expansion velocity. 


The X-ray observation shows that the velocity of the jet-like structure is $\sim 14000-15000 \rm\ km \ s^{-1}$ \citep{2006ApJ...645..283F}. However, due to the projection effect, the true velocity should be higher, and may even reach $\sim 0.1 c$ \citep{2016ApJ...825..102W}. In our model, we consider two velocities for the two structures: $v_1=5000 \rm \ km \ s^{-1}$ for the spherical shock (namely zone 1); and $v_2=30000 \rm \ km \ s^{-1}$ for the jet-like structure (zone 2). The relation between the velocities of the two components is $ v_2=6v_1$.  

For both cosmic ray protons (CRp) and CRe, we assume that their injection energy distributions generally follow the exponential cutoff power law form (ECPL), $\propto E^{-\alpha}\exp(-E/E_{\rm cut})$. The cuttoff energy $E_{\rm cut}$ is the maximum energy that particles can be accelerated. The acceleration timescale for particles in DSA depends on the diffusion of particles in the shock \citep{1983RPPh...46..973D}. Scaled to Bohm limit diffusion, the acceleration timescale of particles with energy $E$ is
\begin{equation}
t_{\rm acc} \approx 1.1\times10^3 ~{\rm yr}\ \eta_{\rm acc} \eta_{\rm g} \frac{E}{1 \rm TeV} \left(\frac{B_u}{1\rm \mu G}\right)^{-1} \left(\frac{v}{10^{8}{\rm cm~ s^{-1}}}\right)^{-2} ,
\end{equation}
where $\eta_{\rm g}\geq1$ accounts for the uncertainty of the particle diffusion relative to Bohm limit, $\eta_{\rm acc}<1$ accounts for the anisotropic scattering due to shock obliquity effect \citep{2008ARA&A..46...89R}, $B_u$ is the upstream magnetic field of the shock front. It should be noted that amplification of magnetic field in the shock can be triggered by the stream instability caused by high energy particles. The magnetic field can reach
$B_{d}^2/8\pi ~\sim 10^{-2}\rho v^{2}$ for downstream region of the shock front \citep{2005A&A...433..229V}, where $\rho$ is the medium mass density, and $B_{u}^2/8\pi \sim 10^{-3}\rho v^{2}$ for upstream region of the shock front \citep{2012A&A...538A..81M}. Hereafter we assume $B_d=\sqrt{10}B_u$ and $B_{2}=(v_{2}/v_{1})B_{1}$ for simplicity, where $B_2$ and $B_1$ denote the magnetic field in zones 2 and 1 respectively.

The particle acceleration suffers from several factors, e.g., the limit of the SNR age, the particle escaping and radiative cooling timescales. For CRe, the synchrotron cooling timescale is given by \citep{1986rpa..book.....R}:
\begin{equation}
\tau_{\rm syn} \approx 1.3\times10^{7}{\rm yr} \left(\frac{E}{1\rm TeV}\right)^{-1} \left(\frac{B_d}{1 \rm \mu G}\right)^{-2} ,
\end{equation}
where $B_d$ is the magnetic filed in the downstream region. The CRe IC cooling timescale is given by \citep{1979rpa..book.....R}:
\begin{equation}
\tau_{\rm IC} = 7.8\times10^5~{\rm yr} \left(\frac{E}{1\rm TeV} \right)^{-1}\left(\frac{u_{\rm ph}}{\rm 1\, eV \ cm^{-3}}\right)^{-1},
\end{equation}
where $u_{\rm ph}$ is the radiation energy density, including the cosmic microwave background ($T=2.7$K, $U_{\rm CMB} \simeq 0.26 \rm \ eV \ cm^{-3}$) and the far infrared background ($T=100$K, $U_{\rm FIR} \simeq 2 \rm \ eV \ cm^{-3}$). The CRe bremsstrahlung cooling timescale is given by \citep{2004Natur.432...75A}:

\begin{equation}
\tau_{\rm bre}\approx4\times10^{7}~{\rm yr}~ \left(\frac{n_{\rm H}}{1\rm cm^{-3}}\right)^{-1}.
\end{equation}
where $n_{\rm H}$ is the number density of background nucleus. Here we take $n_{\rm H}=4 \rm \ cm^{-3}$. For CRp, the pp energy loss timescale is given by \citep{1996A&A...309..917A}:
\begin{equation}
\tau_{\rm pp}\approx6\times10^{7}~{\rm yr}~\left(\frac{n_{\rm H}}{1~{\rm cm}^{-3}}\right)^{-1} . 
\end{equation}

The shock should confine particles efficiently to maintain acceleration, but high energy particles may escape upstream from the shock. Defining the free escaping boundary as a distance $\kappa R$ ahead of the shock front, where $R$ is the SNR shock radius, and $\kappa=0.04-0.1$ \citep{2008ApJ...678..939Z,2011ApJ...731...87E}, the escaping timescale, i.e., the time for particles to cross the boundary, is then given as
\begin{equation}
\tau_{\rm esc}\approx 9\times 10^4 {\rm yr}  \eta_{\rm esc} \eta_{\rm g}^{-1}\left(\frac{E}{1\rm TeV}\right)^{-1} \left(\frac{B_u^f}{1\rm \mu G}\right) \left(\frac{R}{1\rm pc}\right)^2,
\end{equation}
where $\eta_{\rm esc}=\kappa^{2}\eta_{\rm acc}<0.1$, and $B_{u}^{f}$ is the far upstream magnetic field. For Cas A, we take $R=2.5$ pc \citep[][references therein]{2014ApJ...785..130Z}, and $B_{u}^{f} = 5 \rm \ \mu G$, typical for the interstellar medium.

 Finally, the characteristic loss rate of particles with energy $E$ can be given by, for electrons, $1/t_{\rm loss}=1/\tau_{\rm IC}+1/\tau_{\rm syn} + 1/\tau_{\rm bre}+1/\tau_{\rm esc}$, and for protons, $1/t_{\rm loss}=1/\tau_{\rm pp}+ 1/\tau_{\rm esc}$, since the proton radiation energy loss is negligible. The maximum particle energy $E_{\rm cut}$ can be determined by equating $t_{\rm acc}(E_{\rm cut})=\min\{t_{\rm loss}(E_{\rm cut}),t_{\rm age}\}$. For example, Fig. 2 shows all the timescales discussed above for both isotropic and jet-structure shocks, in the case of model B (see Table 1). 

If the particles do not suffer strong energy loss, their energy distribution downstream still follow the ECPL, where the particle number per unity energy $N(E)$ is described as
\begin{equation}
N(E) = A\left(\frac {E}{1\rm TeV} \right)^{-\alpha}\exp\left(-E/E_{\rm cut}\right).
\end{equation}

However, for high energy electrons, they may suffer strong synchrotron and/or IC cooling (Bremsstrahlung cooling is negligible, see Fig 2), then the downstream electron energy distribution deviates from the ECPL, and follow an exponential cutoff broken power law form (ECBPL):
\begin{equation}
N(E) = 
	\begin{cases}
		\ A\left(\frac {E}{1\rm TeV} \right)^{-\alpha}\exp\left(-E/E_{\rm cut}\right) & E \leq E_{b} \\
        \ A\left(\frac {E}{1\rm TeV} \right)^{-\alpha - 1}\frac{E_{b}}{1\rm TeV}\exp\left(-E/E_{\rm cut}\right). & E > E_{b}
	\end{cases}
\end{equation}

 The break energy $E_b$ where a transition of the spectral index from $-\alpha$ to $-\alpha-1$ occurs can be determined by equating $1/t_{\rm age}=1/\tau_{\rm IC}+1/\tau_{\rm syn}$, i.e, $E_{b}=4(100{\rm \mu G}/B_{d})^{2}(350 {\rm yr}/t_{\rm age})$ TeV. In the following modelling of Cas A, we will consider ECPL and ECBPL for proton and electron distributions, respectively. We assume that in the same zone the spectral indices of accelerated electrons and protons are equal, $\alpha_e=\alpha_p$, since the acceleration processes for electrons and protons are expected to be the same. However the spectral indices in different zones may be different since the acceleration process may change in different shocks.

\begin{figure}
\begin{minipage}{0.50\textwidth}
 \includegraphics[width=0.95\textwidth]{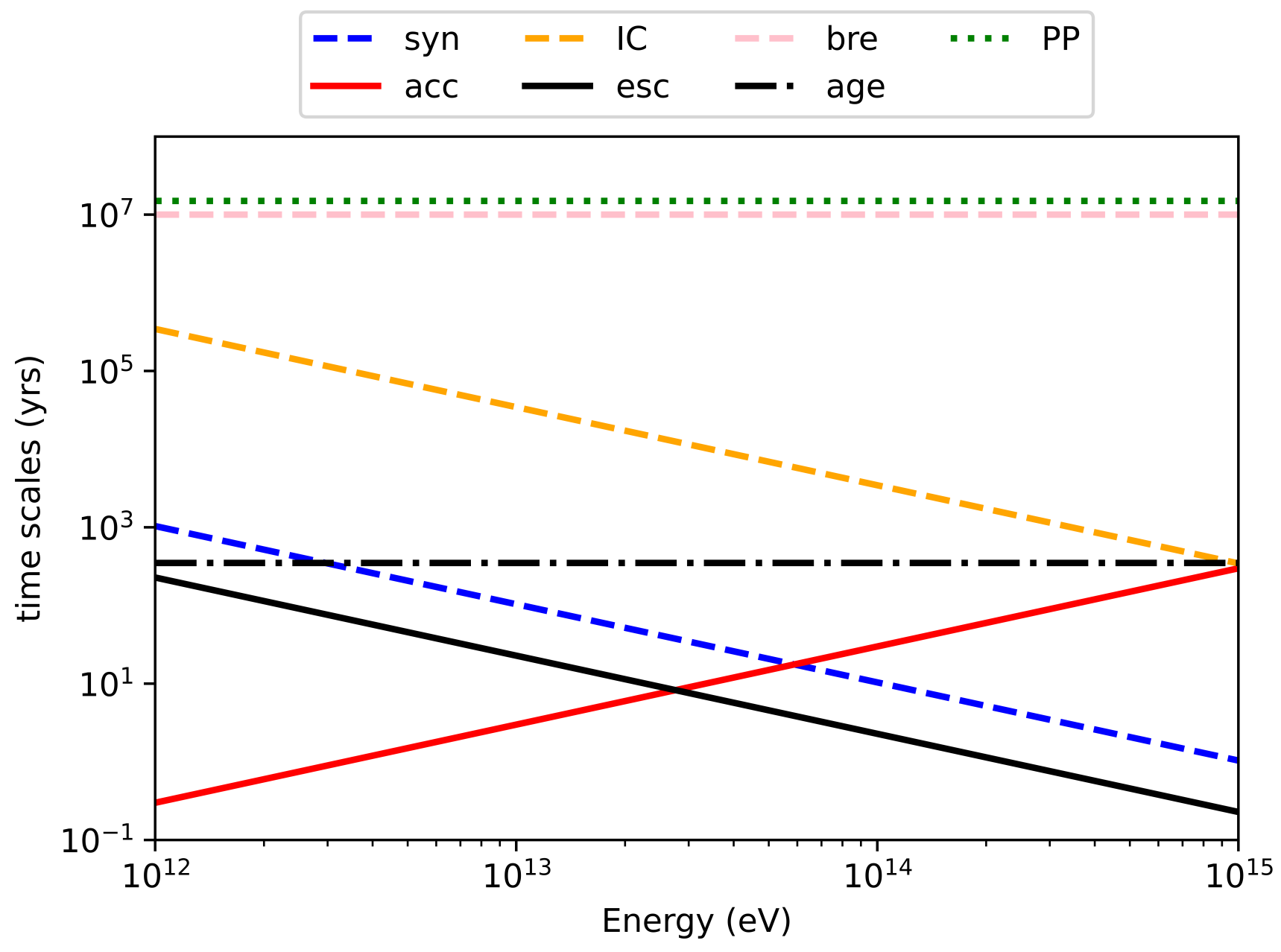}
\end{minipage}
\begin{minipage}{0.50\textwidth}
 \includegraphics[width=0.95\textwidth]{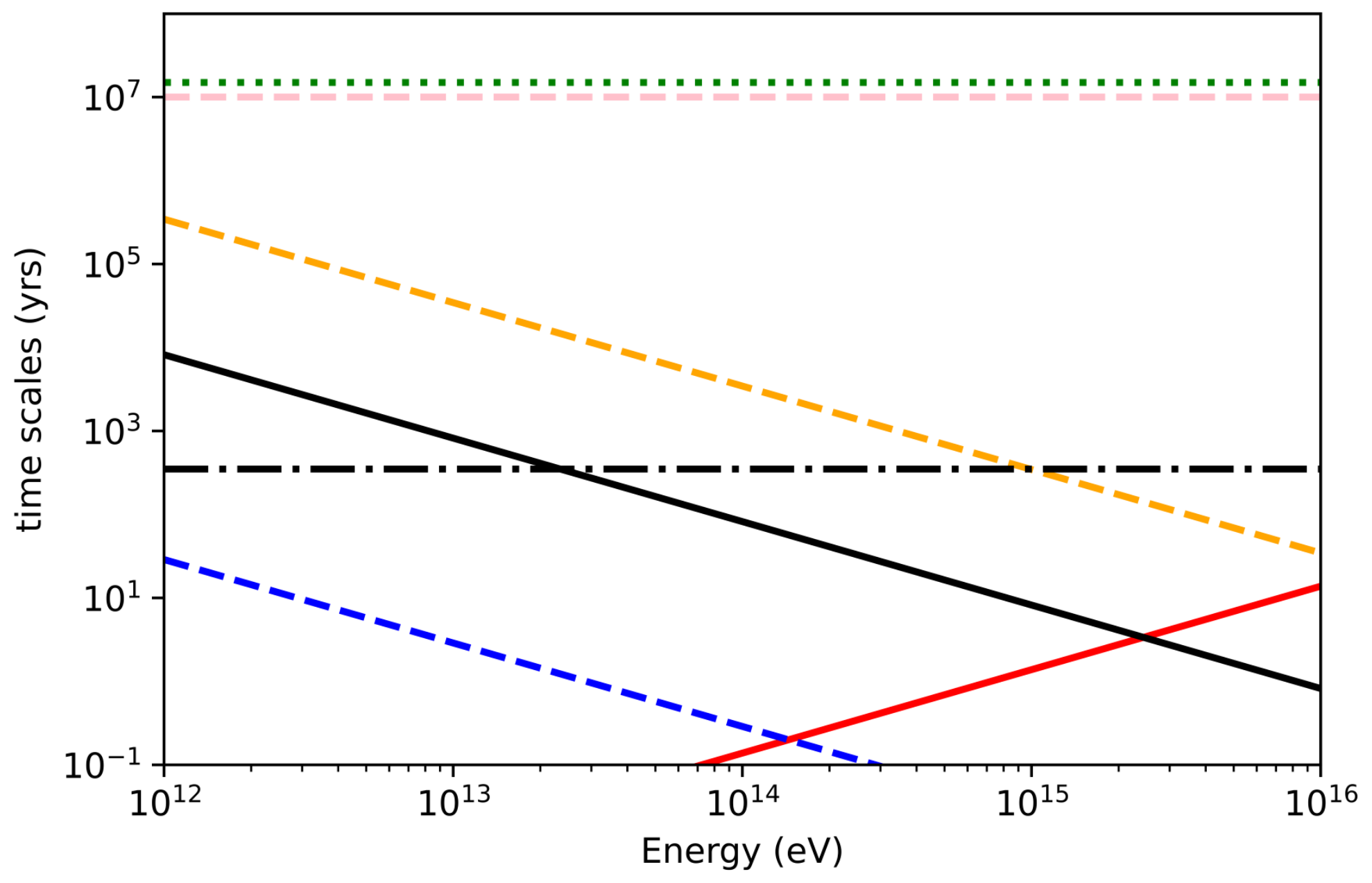}
\end{minipage}
\caption{An example for different timescales in zone 1 ({\em upper panel}) and zone 2 ({\em lower panel}) in model B. Shown are the acceleration (red solid line), escaping (black solid line), IC cooling (orange dashed line), synchrotron cooling (blue dashed line), bremsstrahlung cooling (pink dashed line), pp energy loss (green dotted line) timescales, and the age of Cas A (black chain line). }
\label{fig2} 
\end{figure}

\section{Multiwavelength spectral fitting}

 Based on the model described in Section 2, we fit the observed multi-band data using the public code $naima$\footnote{https://github.com/zblz/naima}\citep{2015ICRC...34..760K}. We will adopt three models in the fitting: Model A is a simple one-zone model for the emission region, i.e., zone 1 in Fig 1; Model B is the two-zone model for the asymmetrical expansion, where the particle spectral indices and the electron to proton energy ratios $W_e/W_p$ ($W_e=\int_{10\rm MeV}^{\infty} EN(E)dE$, and $W_p=\int_{1 \rm GeV}^{\infty} EN(E)dE$ are the electron and proton energies in the shock, respectively) in the two zones are the same; and Model C is same as Model B but allows the particle spectral index and electron to proton energy ratio to change between two zones. We take $\eta_{\rm acc}=0.11$, $\eta_{\rm g}=2.2$, and $\eta_{\rm esc}=1.8 \times 10^{-4}$ in the fitting. The parameter values in the fitting are presented in Table 1 for the three models. We will neglect the secondaries from pp collisions contributing to the multi-band radiation in Cas A (as evidenced in the Appendix).

In Model A, we find that the downstream magnetic field is well constrained. On the one hand, the soft X-ray data requires that $E_{b}$ can not be too low, leading to an upper bound on $B_{d}$; on the other hand, the magnetic field cannot be too low so that the IC emission would overshoot the TeV $\gamma$-ray data, which leads to a lower bound on $B_{d}$. These two competing factors indeed give a very narrow range of $B_d$, and finally we get, from the fitting (see Fig. 3), $B_{d}\approx110\rm \mu G$ in zone 1 (see Table 1). 

However, it should be pointed out that, as obviously shown in Fig. 3, Model A can well account for the data from radio to GeV-TeV range, except for the hard X-rays. The synchrotron radiation can account for the data up to $\sim 60$ keV and cannot well fit the hard X-rays at $\sim 60 -220$ keV. The excess in the hard X-rays may require extra contribution in addition to the isotropic expanding shell.

Next, in Model B we consider the extra contribution of a jet-like structure (zone 2). The parameters for zone 2 are sensitively dependent of the hard X-ray data from $\sim$ 60--220 keV obtained by INTEGRAL-IBIS. The multi-band spectral fitting result of Model B is presented in Fig. 4. Comparing the results in Figs. 3 and 4, the emission from the jet-like structure (zone 2) in Model B can well explain the hard X-ray data in $\sim 60 -220$ keV (see the residuals in Figs. 3 and 4). 

\begin{table*}
\scriptsize
\caption{The parameter values for the multi-band spectrum fitting of Cas A.}
\begin{center}
\begin{tabular}{c c c c c c c c c c c c}
\hline \hline
Model & zone& $B_{d}(\mu G)$ & $A_{e}$($\rm TeV^{-1}$) & $A_{p}$($\rm TeV^{-1}$) & $\alpha_{e}$ & $\alpha_{p}$ & $E_{ecut}$(TeV) & $E_{pcut}$(TeV) & $W_{e}$(erg) & $W_{p}$(erg)  & $E_{ b}$(TeV) \\
\hline
A & & 112 & $4.8\times10^{46}$ & $3.0\times10^{48}$ & 2.47 & 2.47 & 25.0 & 27.6 & $3.7\times10^{49}$ & $2.0\times10^{50}$  & 3.2\\
\hline
B & 1 & 112& $4.5\times10^{46}$ & $3.0\times10^{48}$ & 2.47 & 2.47 & 25.0 & 27.6 & $3.4\times10^{49}$ & $2.0\times10^{50}$  & 3.2\\
 &  2& 672 & $5.0\times10^{44}$ & $3.3\times10^{46}$ & 2.47 & 2.47 & 144 & 2440 & $3.8\times10^{47}$ & $2.2\times10^{48}$  & 0.09\\
\hline
C& 1 & 123& $4.1\times10^{46}$ & $3.0\times10^{48}$ & 2.47 & 2.47 & 25.7 & 28.6 & $3.1\times10^{49}$ & $2.0\times10^{50}$   & 2.6\\
 &2 & 738& $1.3\times10^{44}$ & $1.4\times10^{47}$ & 2.1 & 2.1 & 137 & 2520 & $5.2\times10^{45}$ & $2.5\times10^{48}$  & 0.07\\
\hline
\end{tabular}
\end{center}
\label{table1}
\end{table*}

\begin{figure}
\begin{minipage}{0.50\textwidth}
 \includegraphics[width=0.88\textwidth]{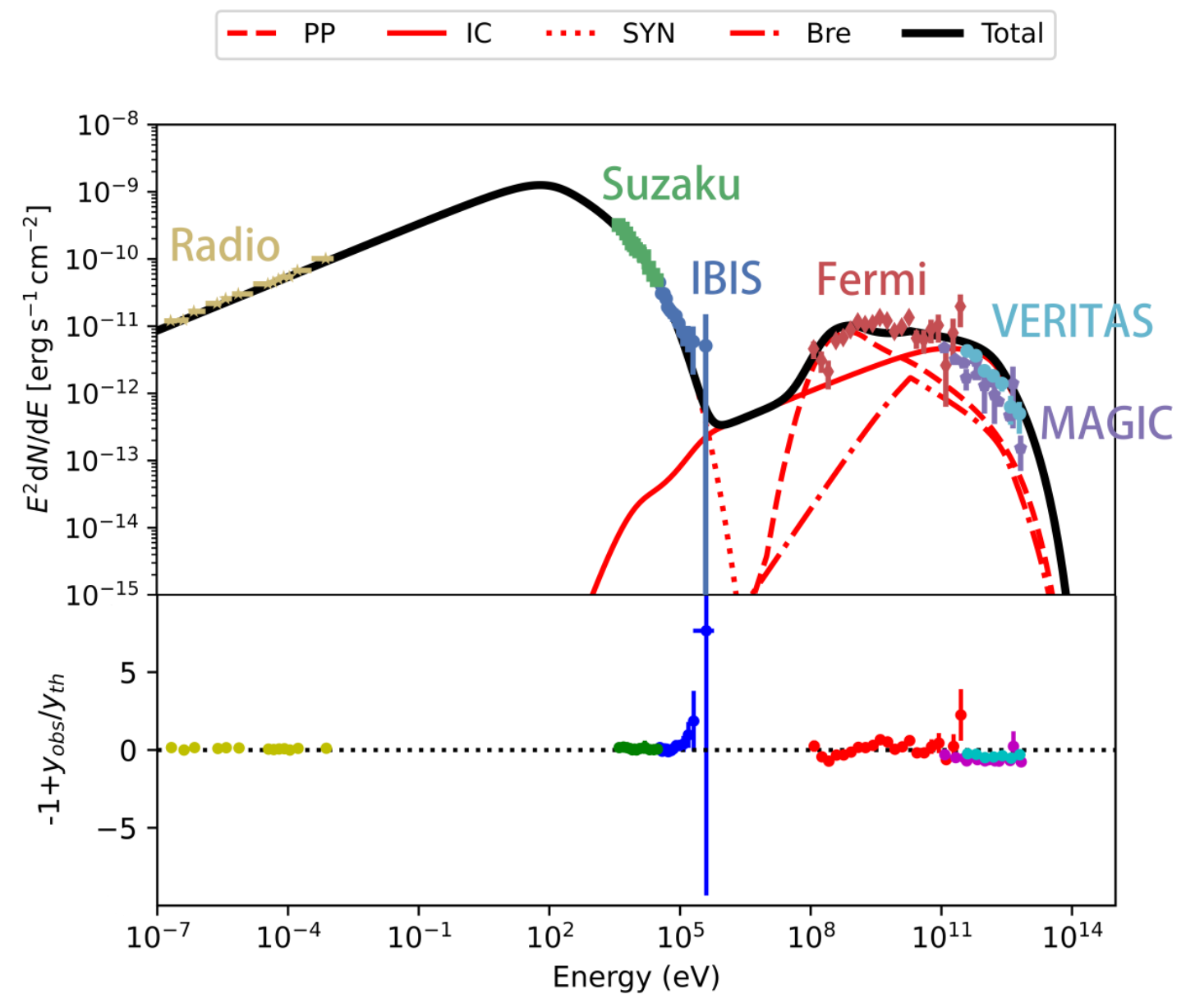}
  \includegraphics[width=0.88\textwidth]{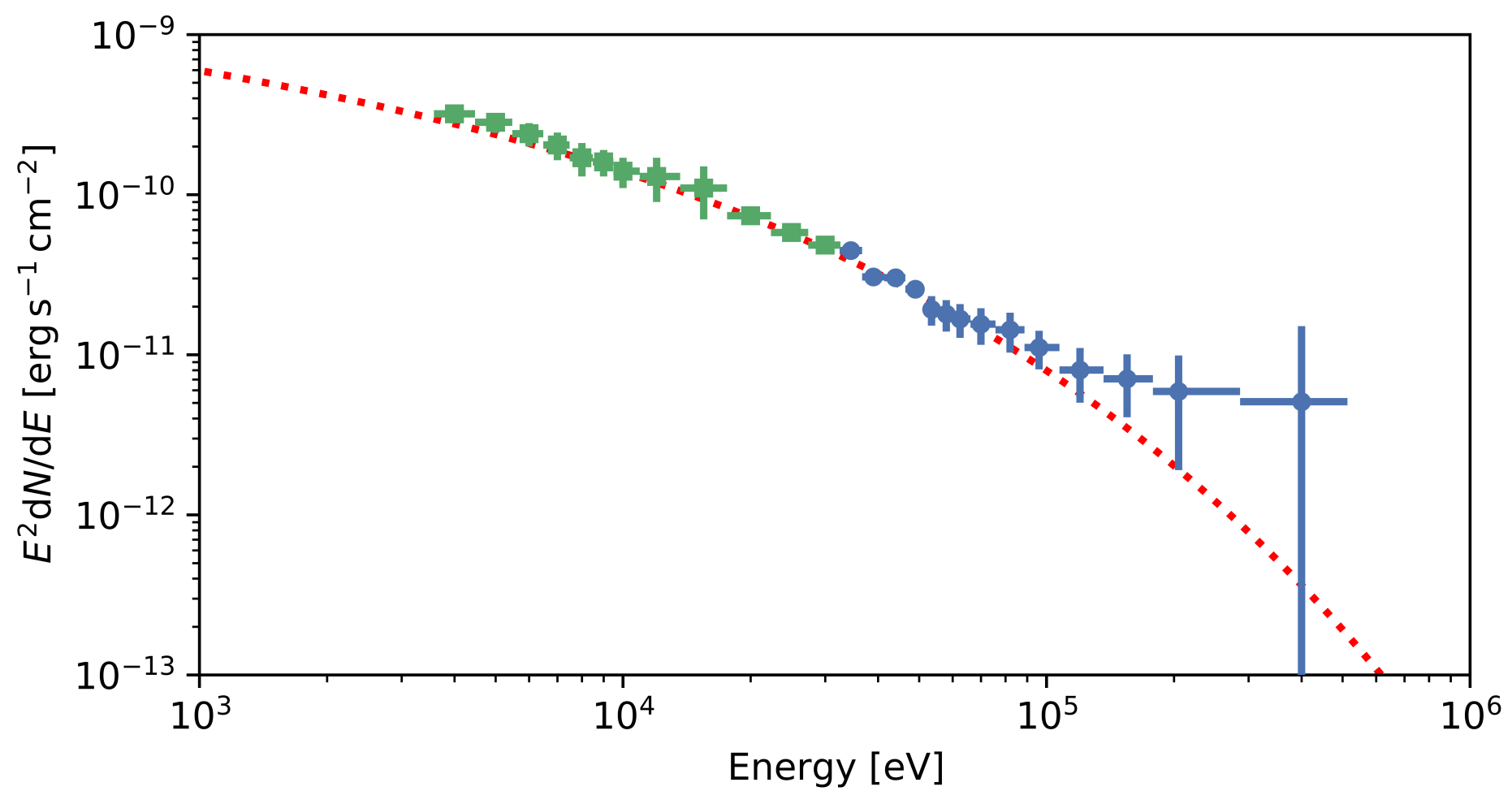}
\end{minipage}
\caption{The spectral fitting of the multi-band data for Cas A in Model A (the one-zone model). The {\em upper panel} is the broadband fitting, the {\em middle panel} is the fitting residuals, and the {\em bottom panel} shows the zoom-in of the X-ray range. The radio data are taken from \protect\cite{2014ARep...58..626V}, X-ray data from INTEGRAL-IBIS \citep{2016ApJ...825..102W} and Suzaku \citep{2009PASJ...61.1217M}, and $\gamma$ data from Fermi-LAT \citep{2010ApJ...710L..92A}, VERITAS \citep{2008AIPC.1085..357H,2020ApJ...894...51A}, and MAGIC \citep{2017MNRAS.472.2956A}. }
\label{fig3}
\end{figure}

\begin{figure}
\begin{minipage}{0.50\textwidth}
 \includegraphics[width=0.88\textwidth]{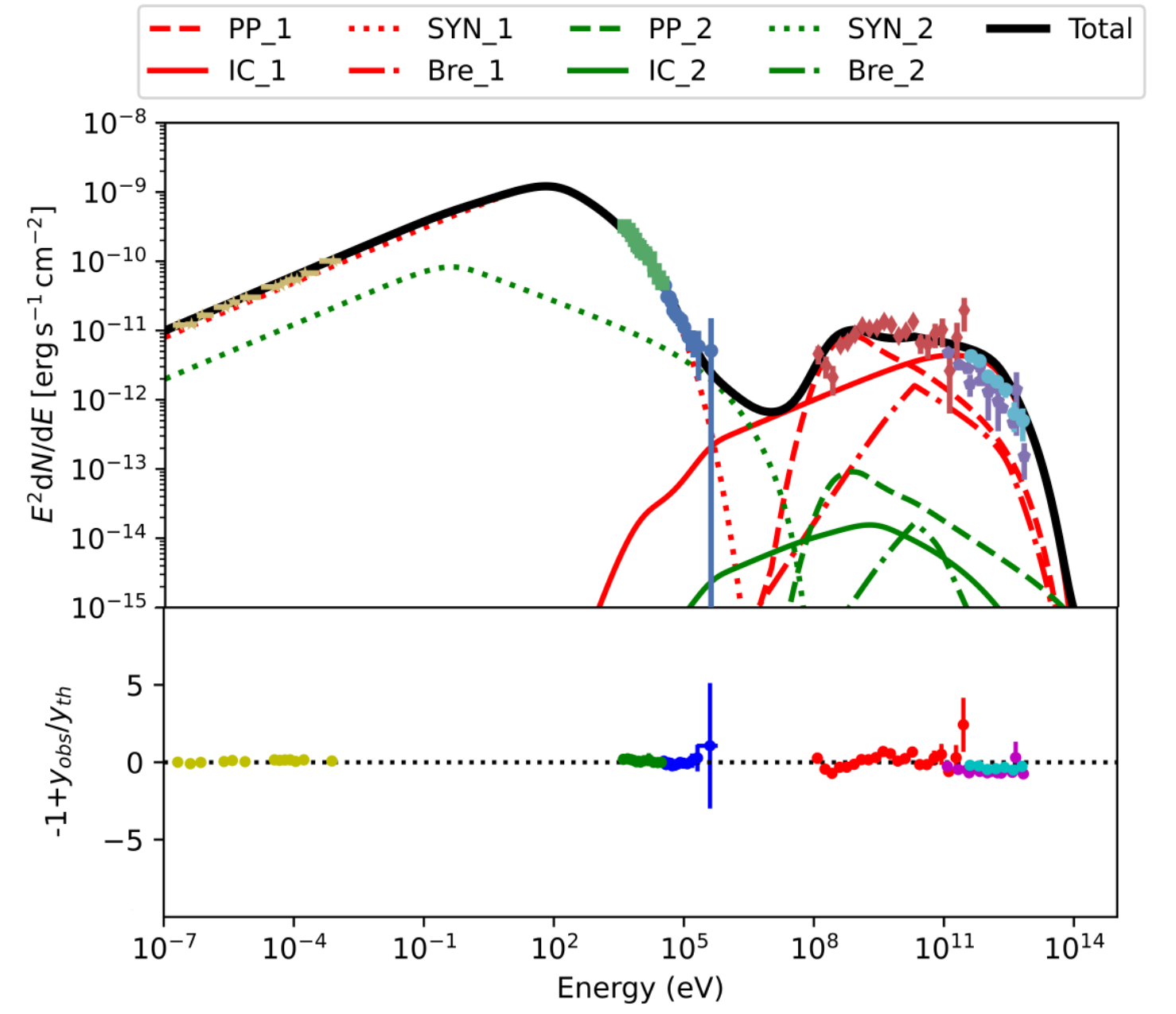}
  \includegraphics[width=0.88\textwidth]{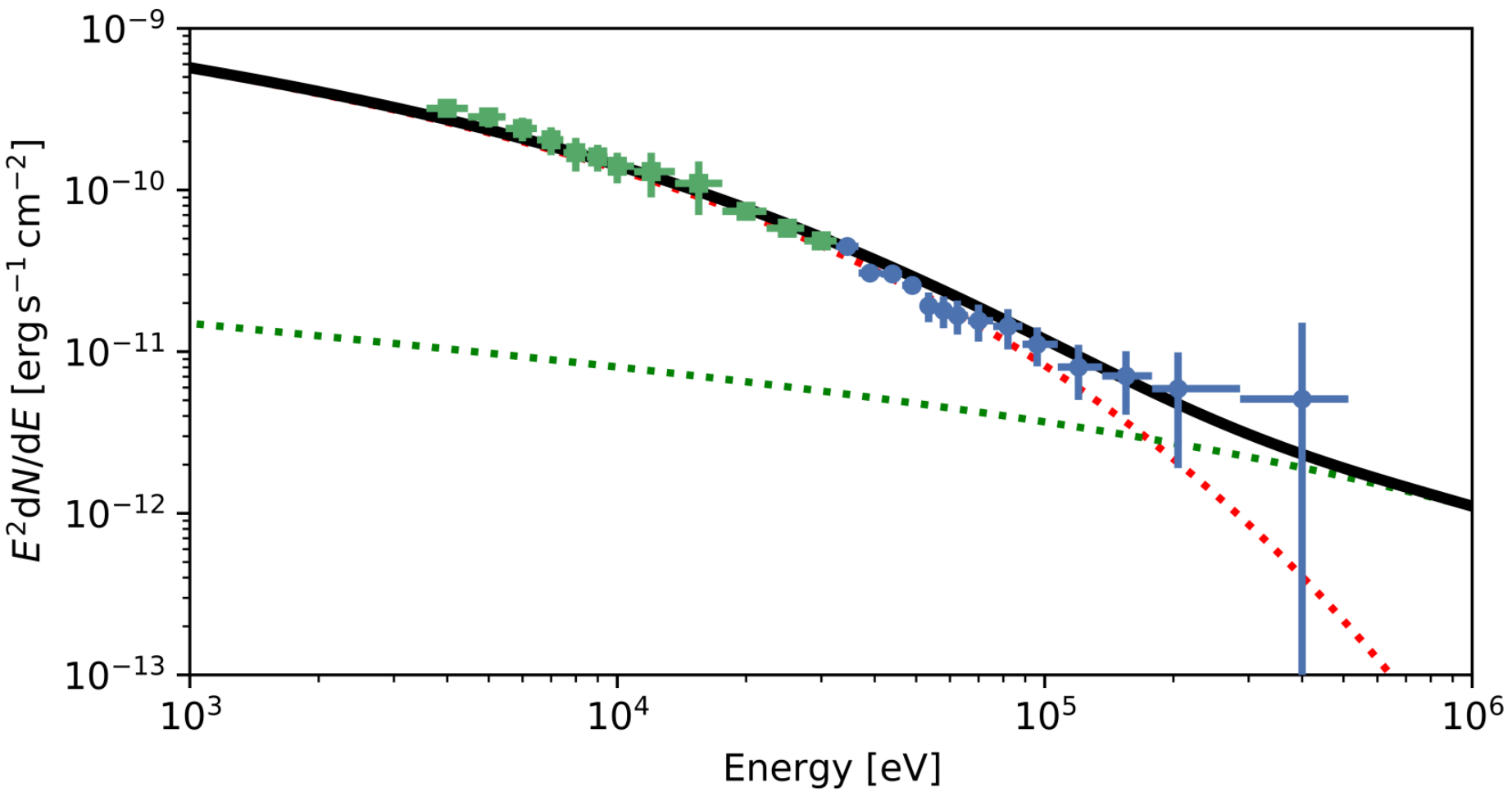}
\end{minipage}
\caption{Same as Fig 3 but for Model B. The red lines present contribution from zone 1, and green lines from zone 2. }
\label{fig4}
\end{figure}

As shown in Table 1, the non-thermal particle energy in zone 2 is much smaller than that in zone 1. The total energy of accelerated electrons and protons in the shock is $\sim 2.0 \times 10 ^{50} \rm \ erg$. If the shock converts a fraction 0.1 of the total kinetic energy into the accelerated particles, the total kinetic energy is $\sim 2\times 10^{51}$ erg for Cas A, which is about twice as large as the conclusion of \cite{2003A&A...398.1021W}.

For the $\gamma$-ray emission, we find that given the magnetic filed fixed at $B_{d}=112\rm \mu G$, the hadronic process in zone 1 dominates the radiation from $10^{8}$ eV to $\sim 10^{10}$ eV, but the leptonic process in zone 1 becomes dominant above $10^{10}$ eV. It should be pointed out that the dominant emission component in the GeV- TeV band is produced in zone 1. Thus, the contribution of the protons in zone 2 to the GeV - TeV radiation is very small, and it is difficult to constrain the parameters for protons in zone 2 in Model B. 

Finally in Model C, since the isotropic shell and the fast jet structure may work differently in particle acceleration, the two zones may not have the same spectral index and $W_e/W_p$ ratio. A fitting result with harder spectral index in zone 2, $\alpha=2.1$ in Model C is shown in Fig. 5 and the parameter values are presented in Table 1.

\begin{figure}
\begin{minipage}{0.50\textwidth}
 \includegraphics[width=0.88\textwidth]{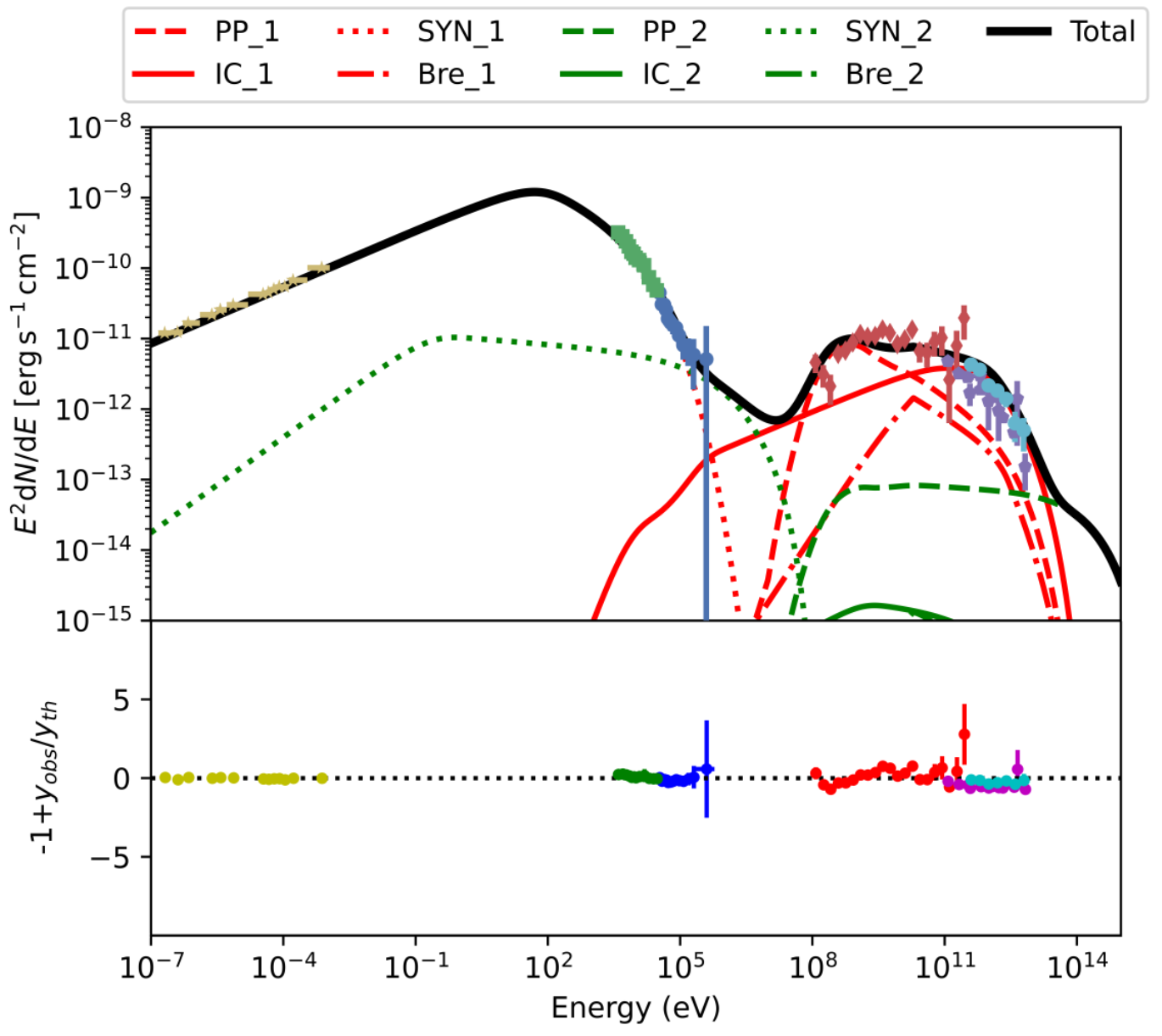}
 \includegraphics[width=0.88\textwidth]{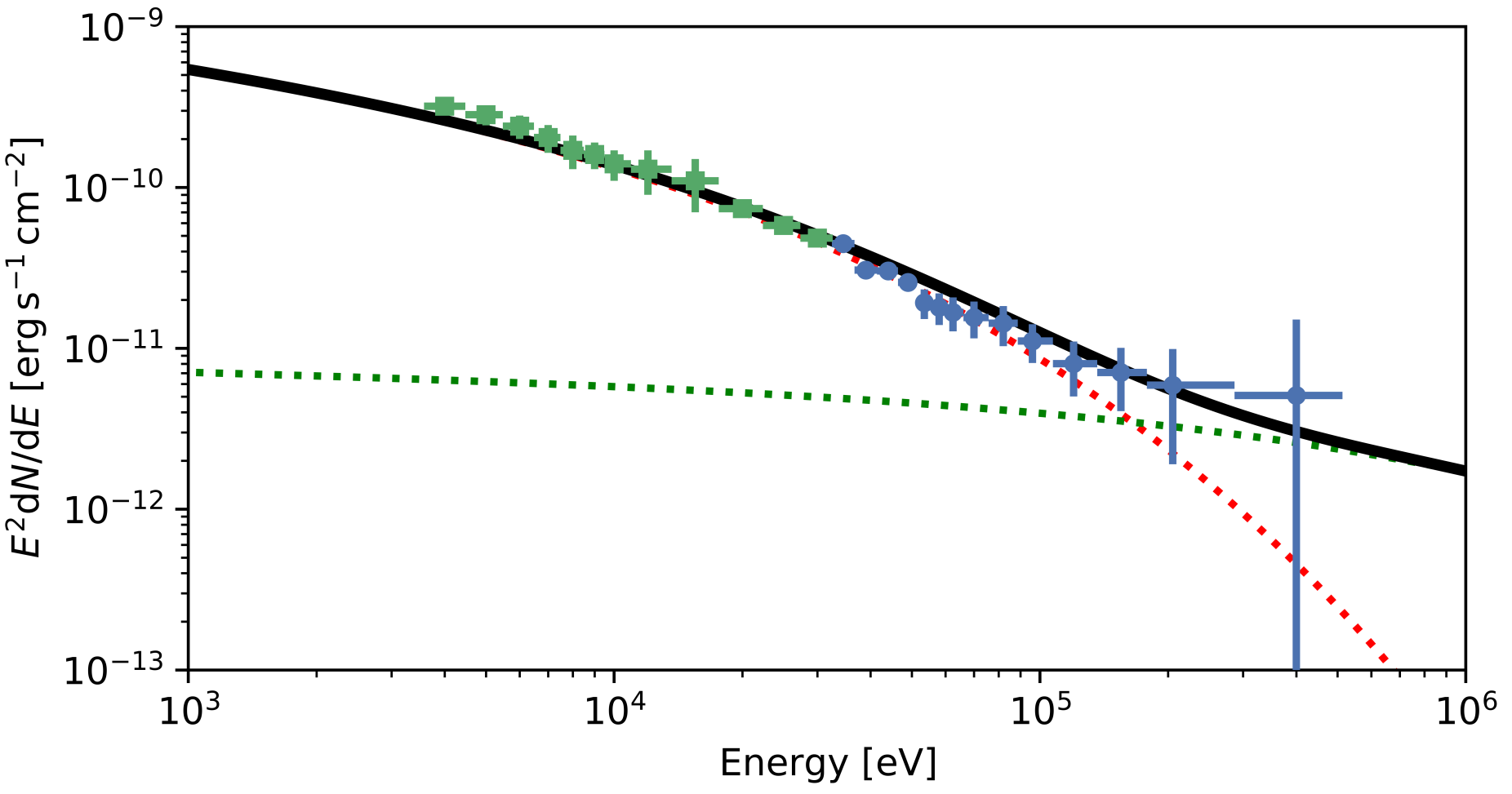}
\end{minipage}
\caption{ Same as Fig. 4 but for Model C where the particle injection spectral index differs between zone 1 and 2.}
\label{fig6}
\end{figure}

From Fig. 5, we can find that the jet component may produce a gamma-ray flux at the level of $\sim 10^{-13} \rm erg\ cm^{-2}\ s^{-1}$, which, although has a negligible contribution to the GeV-TeV emission observed by Fermi and VERITAS, dominates at $\ga$100 TeV. This is encouraging future LHAASO and CTA observations to test whether the two shocks in zones 1 and 2 work differently, and whether Cas A can be a PeVatron in the Galaxy.

\section{Summary and conclusions}

The physical origin of the non-thermal high energy radiation of Cas A is still under debate. We proposed an asymmetrical model based on the multi-band observations, consisting of a fast jet-like structure and a slow, isotropically expanding shell. Our model can well explain the multi-wavelength data of Cas A from radio to gamma-ray bands. 

The synchrotron radiation from primary electrons is the main contribution to the radio to hard X-ray emission in Cas A. The radiation from the isotropic component exhibits a spectral deviation at $\sim$ 60 keV. The spectral fitting considering the cooling effect of electrons put strong constraint on the magnetic field, resulting in $B_d \sim 110 \mu$G in zone 1. Previous observations suggest that the hard X-ray emission is dominated by some bright knots with a different spectral index from the outer filaments \citep{2015ApJ...802...15G}, which is consistent with Model C in that the two zones have different spectral indices. In addition, we note that if the velocity of jet-like structure is high enough ($\sim 0.1$ c), the jet component in Model C may generate sub-PeV gamma-rays with a flux of $\sim 10^{-13} ~{\rm erg\ cm^{-2}\ s^{-1}}$ around 100 TeV which could be detected by LHAASO and CTA. 

We argue that one-zone model fails in fitting the hard X-rays in $\sim$ 60--220 keV, which could be contributed by the jet-like component.  Alternatively, the hard X-rays could be explained by the reverse shock \citep{2019ApJ...874...98Z}, but different from our fitting results, the main contribution in hard X-ray bands of $60 - 220$ keV is bremsstrahlung process in the reverse shock assumption while our model indicates that the main contribution in hard X-rays is the synchrotron process of zone 2. 

 Both Model B and C work well in fitting the GeV-TeV data of Fermi, VERITAS and MAGIC, however the particle acceleration works differently between them. As shown in Table 1, the $W_e/W_p$ ratio and the particle spectra are set to be the same for zones 1 and 2 in Model B, but be flexible in Model C. Furthermore, Model B predicts a low flux, $< 10^{-14} ~{\rm erg\ cm^{-2}\ s^{-1}}$ around 100 TeV, but Model C suggests that there may be a gamma-ray flux of $\sim 10^{-13} ~{\rm erg\ cm^{-2}\ s^{-1}}$ around 100 TeV, which can be reached by LHAASO's sensitivity \citep{2021Natur.594...33C}. The particle acceleration is poorly understood. Given the different physical condition in the isotropic and jet-like structure components, especially the very different shock velocities, the particle acceleration process may work differently. Future 100 TeV observations are helpful to resolve the two models, and investigate the particle acceleration physics.


The jet-like structure produces a soft X-ray flux much smaller than that from the isotropic component, and may be too dim to be detected by Chandra. The eRosita will observe X-rays up to 30 keV, with the sensitivity $\sim 7\times 10^{-13} \rm erg\ cm^{-1}\ s^{-1}$ and angular resolution $\sim 10"$ from 2-8 keV, and $\sim 10^{-11} \rm erg \ cm^{-1}\ s^{-1}$ and $\sim 45"$ from 8-30 keV \citep{2021A&A...647A...1P,2021A&A...656A.132S}. However, the jet structure produces an X-ray flux below the sensitivity of eRosita from 8-30 keV, so that eRosita would be difficult to distinguish the jet structure. Moreover, one does not expect to see filaments associated with the jet produced shock, unless the light of sight is along the shock surface. 

The maximum energy of particles accelerated in jet-like structure is up to one hundred TeV for electrons, and around PeV for protons. However, it is possible that the jet may only consist of fast-moving knots, which generate bow shocks. According to \cite{2006ApJ...645..283F}, the knots can still produce electrons that account for the X-ray data in our model. If the particles can be well confined in the bow shock and get accelerated, then the available acceleration time is the dynamical time of the knots, $\sim R/V_{\rm knot}$, where $R\sim 0.1-1$ pc is the distance of the knots away from the explosion center. In this case, the maximum energy of accelerated particles is same as a jet, $E_{\rm max} \sim 1.2{\rm PeV} (R/0.1 {\rm pc}) (V_{\rm knot}/0.1c) (B_{d}/600\mu{\rm G})$, so these fast-moving knots could accelerate protons to the PeV range.

In summary, the asymmetrical model with jet-like structure can well fit the hard X-ray spectrum from $\sim$ 60 keV to 220 keV. The current observations and model fittings still cannot confirm whether Cas A is a PeVatron or not. The present and future instruments, e.g., LHAASO and CTA, are encouraging to solve the PeVatron mystery in the Galaxy \citep{2021Natur.594...33C}. 

\section*{Acknowledgments}
 We are grateful to the referee for the helpful comments. This work is supported by the National Key Research and Development Program of China (Grants No. 2021YFA0718500, 2021YFA0718503) and the NSFC (12133007, U1838103, 11622326, 11773008, 11703022, 11773003, and U1931201), the Fundamental Research Funds for the Central Universities (No. 2042021kf0224), and the China Manned Space Project (CMS-CSST-2021-B11).

\section*{Data Availability}
The data used in this paper were collected from the previous literatures. These data of radio, X-ray and gamma-ray observations are public for all researchers. The codes used in this work like $naima$ and $cparamlib$ are also public in the available websites.

\bibliographystyle{mnras}
\bibliography{references}


\appendix

\section{ESTIMATION OF SECONDARY ELECTRON RADIATION}
In order to study the possible contribution by radiation of the secondary electrons  (including positrons) from pp collisions, we consider an extreme case that all the $\gamma$-rays from Cas A are produced by pp collisions. The jet-like component is also considered. The spectral fitting results in the GeV-TeV band with only hadronic processes are shown in Fig. A1 and Table A1.

\begin{table*}
\scriptsize
\caption{The fitted parameters of pp collisions.}
\begin{center}
\begin{tabular}{c c c c c}
\hline \hline
 zone  & $A_{p}$($\rm TeV^{-1}$) & $\alpha_{p}$  & $E_{pcut}$(TeV) & $W_{p}$(erg) \\
\hline
 1 & $1.5\times10^{49}$ & 2.1 & 9.0 & $2.0\times10^{50}$ \\
 2& $1.3\times10^{47}$  & 2.1 & 1944.0 & $2.3\times10^{48}$ \\
\hline
\end{tabular}
\end{center}
\label{tableA1}
\end{table*}

\begin{figure}
\begin{minipage}{0.50\textwidth}
 \includegraphics[width=0.88\textwidth]{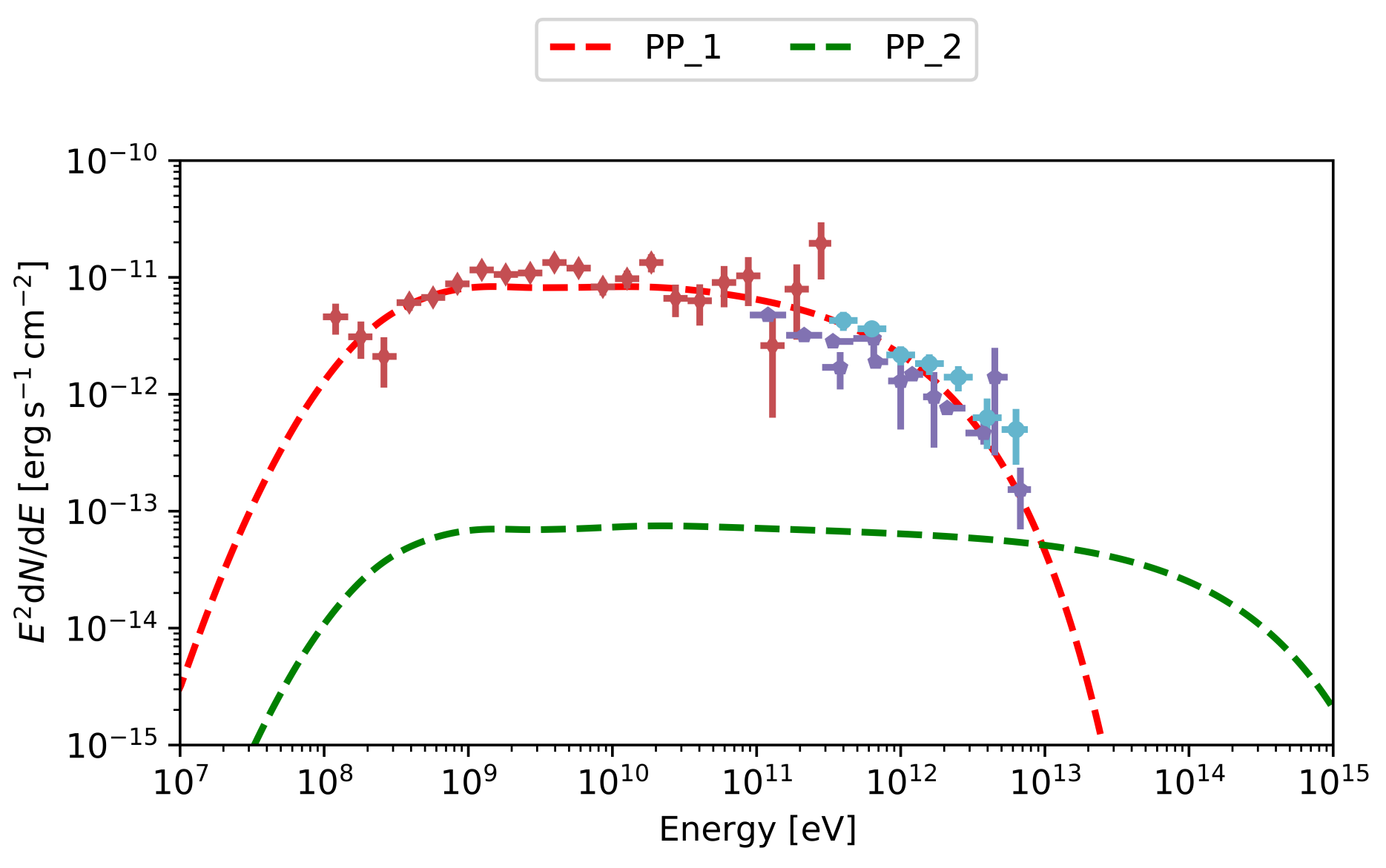}
\end{minipage}
\caption{The spectral fitting of the GeV-TeV gamma-ray data considering only pp collisions. The red and green lines corresponds to contribution from zone 1 and 2, respectively.}
\label{figA1}
\end{figure}

To calculate the products of secondary electrons from pp collisons, we adopt the cparamlib package\footnote{https://github.com/niklask/cparamlib}\citep{2006ApJ...647..692K}. The spectrum of secondary electrons should show a break due to the cooling effect\citep{2011hea..book.....L}. Because the synchrotron cooling is much stronger than the other cooling processes for electrons in our case, we only consider the synthrotron cooling here. Since the synchrotron radiation of secondary electrons are strongly influenced by the magnetic field, we consider two cases: we take $B_{d}=200\rm \mu G$ and $B_{d}=1200\rm \ \mu G$ for zone 1 and 2 respectively in Model A1; and $1000\rm \mu G$ and $6000\rm \mu G$ in Model A2, very extreme in SNR shocks. The energy spectra of the secondary electrons and their synchrotron radiation are shown in Fig. A2.  We can see that the radiation of secondary electrons is very low compared with the observed X-ray flux, and has no influence on our multi-wavelength fitting results.

\begin{figure}
\begin{minipage}{0.50\textwidth}
 \includegraphics[width=0.88\textwidth]{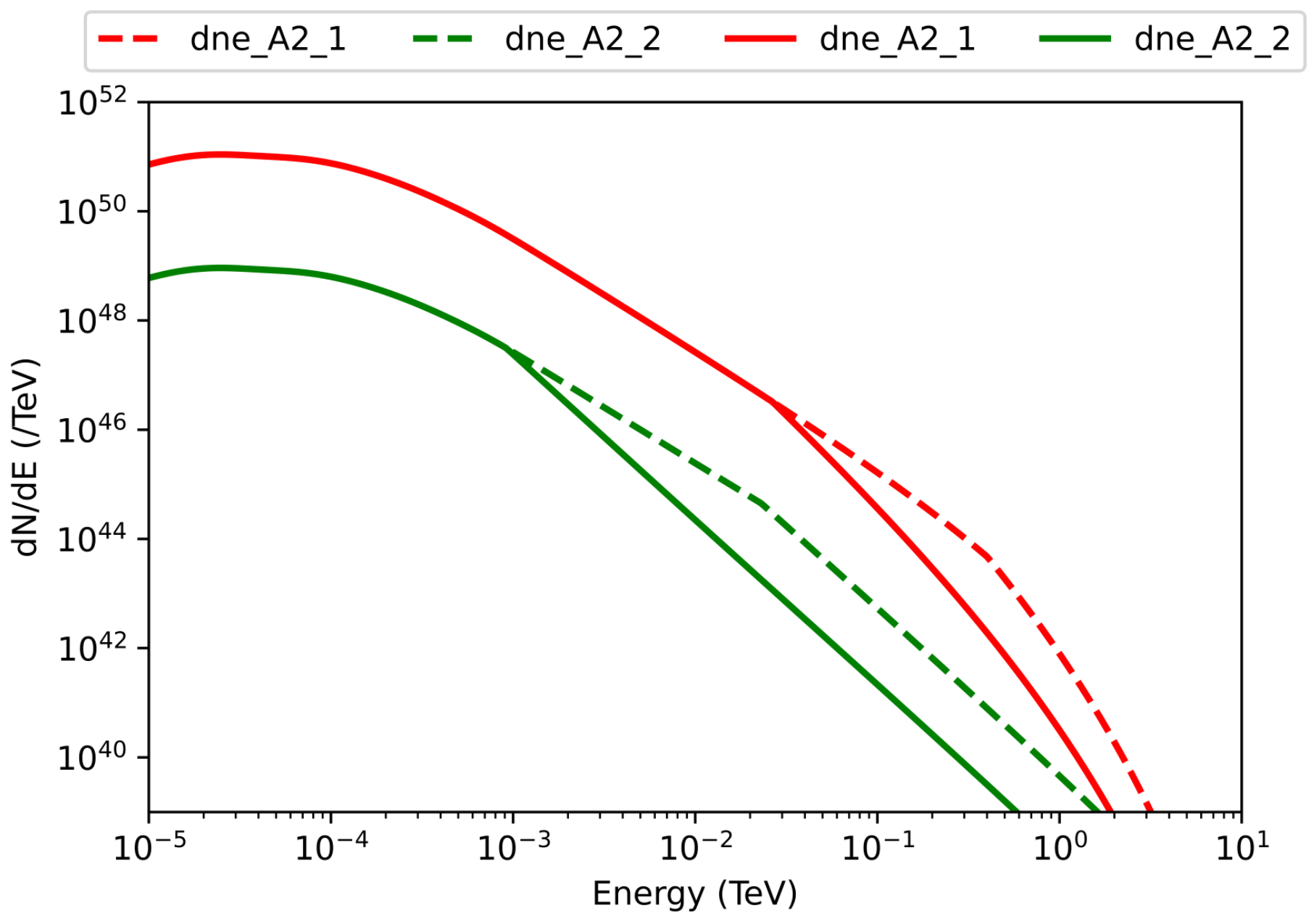}
\end{minipage}
\begin{minipage}{0.50\textwidth}
 \includegraphics[width=0.88\textwidth]{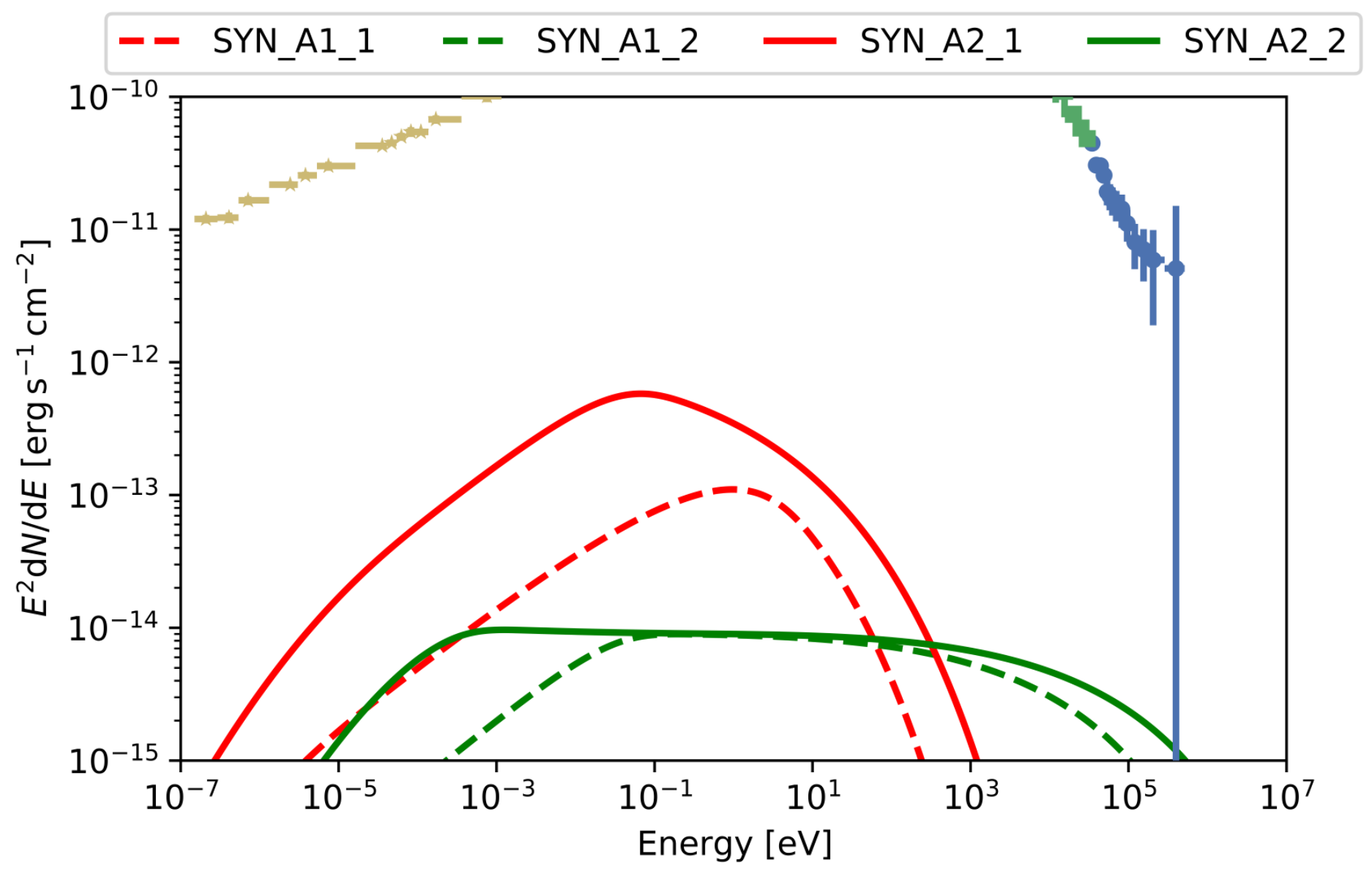}
\end{minipage}
\caption{The secondary electron energy distibution (upper panel) and their synchrotron radiation spectrum in comparison with the X-ray data (lower panel). Dashed and solid lines represent model A1 and A2, respectively. Red and green lines represent the radiation from zone 1 and 2, respectively.}
\label{figA2}
\end{figure}



\label{lastpage}

\end{document}